\documentclass[preprint,showpacs,preprintnumbers,amsmath,amssymb]{revtex4}

\usepackage{graphicx}
\usepackage{epsfig}		
\usepackage{dcolumn}
\usepackage{bm}

\def\0{\mbox{\tiny $0$}}
\def\1{\mbox{\tiny $1$}}
\def\2{\mbox{\tiny $2$}}
\def\3{\mbox{\tiny $3$}}
\def\4{\mbox{\tiny $4$}}
\def\5{\mbox{\tiny $5$}}
\def\6{\mbox{\tiny $6$}}
\def\7{\mbox{\tiny $7$}}
\def\8{\mbox{\tiny $8$}}
\def\9{\mbox{\tiny $9$}}

\def\bb#1{\mbox{\footnotesize $(#1)$}}

\def\x{\mbox{\tiny $x$}}
\def\ii{\mbox{\tiny $i$}}
\def\mi{\mbox{\tiny $-$}}

\def\pmp{\mbox{\tiny $\mp$}}

\begin{document}
\title{Relativistic tunneling and accelerated transmission}

\author{A. E. Bernardini}
\email{alexeb@ifi.unicamp.br}
\affiliation{Instituto de F\'{\i}sica Gleb Wataghin, UNICAMP,
PO Box 6165, 13083-970, Campinas, SP, Brasil.}

\date{\today}

\begin{abstract}
We obtain the solutions for the tunneling zone of a one-dimensional electrostatic potential in the relativistic (Dirac to Klein-Gordon) wave equation regime when the incoming wave packet exhibits the possibility of being almost totally transmitted through the potential barrier.
The conditions for the occurrence of accelerated and, eventually, superluminal tunneling transmission probabilities are all quantified and the problematic superluminal interpretation originated from the study based on non-relativistic dynamics of tunneling is reevaluated.
The treatment of the problem suggests revealing insights into condensed-matter experiments using electrostatic barriers in single- and bi-layer graphene, for which the accelerated tunneling effect deserves a more careful investigation.
\end{abstract}

\date{\today}
\maketitle

Finding a definitive interpretation for the nature of {\em superluminal} barrier tunneling has brought up a fruitful discussion in the literature \cite{Hau89,But03,Win03A,Win03,Olk04} since pulses of light and microwaves appear to tunnel through a barrier at speeds faster than a reference pulse moves through a vacuum \cite{Nim92,Ste93,Spi94,Hay01}.
Tunneling occurs when a wave impinges on a thin barrier of opaque material and some small amount of the wave {\em leaks} through to the other side.
The superluminal experiments that promoted the controversial discussions were performed with a lattice of layers of transparent and opaque materials arranged so that waves of some frequencies are reflected (through destructive interference) but other frequencies pass through the lattices in a kind of {\em filter} effect correlated to the Hartman effect \cite{Har62}.

In all cases described by the non-relativistic (Schroedinger) dynamics \cite{Olk04}, the pulse (wave packet) that emerges from the tunneling process is greatly attenuated and front-loaded due to the {\em filter} effect (only the leading edge of the incident wave packet survives the tunneling process without being severally attenuated to the point that it cannot be detected).
If one measures the speed by the peak of the pulse, it looks faster than the incident wave packet.
Since the transmission probability depends analytically on the momentum component $k$  ($T \equiv T\bb{k}$), the initial (incident wave) momentum distribution can be completely distorted by the presence of the barrier of potential.
As there is no sharp beginning to a pulse, we cannot declare the instant of its arrival at a certain point.
Thus the computation of the tunneling time becomes fundamentally meaningless.
We could only watch the rising edge of the pulse and try to recognize what is arriving \cite{But03}.

By employing a tunneling dynamics described by a relativistic wave equation we can reevaluate the most part of these difficulties.
In this scenario, the existence of formal analogies between the barrier tunneling of the pulses of
light and the tunneling transmission of relativistic particles allows for a close correspondence between quantum relativistic motions described by the Klein-Gordon equation and electromagnetic wave propagations in the presence of dissipation \cite{Ran95}.
Here we demonstrate with complete mathematical accuracy that, in some limiting cases of the relativistic (Klein) tunneling phenomena where the relativistic kinetic energy is approximately equal to the potential energy of the barrier, and $m c L /\hbar << 1$, particles with mass $m$ can pass through a potential barrier $V_{\0}$ of width $L$ with transmission probability $T$ approximately equal to one.
Since $T \sim 1$, the analytical conditions for the stationary phase principle applicability which determines the tunneling (phase) time for the transmitted wave packets are totally recovered.
Differently from the previous (non-relativistic) tunneling analysis, the original momentum distribution is kept undistorted and there is no {\em filter} effect.
The tunneling time is then computed for a completely undistorted transmitted wave packet, which legitimizes any eventual accelerated transmission.

Some authors consider difficult and perhaps confusing the treatment of all interactions of plane waves or wave packets with a barrier potential using a relativistic wave equation \cite{Del03,Cal99,Dom99,Che02}.
This is because the physical content depends upon the relation between the barrier height $V_{\0}$ and the mass $m$ of the incoming (particle) wave, beside of its total energy $E$.
In the first attempt to evaluate this problem, Klein \cite{Kle29} considered the reflection and transmission of electrons of  incidence energy $E$ on the potential step $V\bb{x} = \Theta\bb{x}V_{\0}$ in the $2$-dimensional time-independent Dirac equation which can be represented in terms of the usual Pauli matrices \cite{Zub80} by\footnote{$\Theta\bb{x}$ is the Heavyside function.}
\small\begin{equation}
\left[\sigma^{3}\sigma^{\ii}\partial_{\ii} - (E - \Theta\bb{x_{\1}}V_{\0}) - \sigma^{3} m\right]\phi\bb{k, x_{\1},x_{\2}} = 0,
~~(\mbox{from this point}~ c = \hbar = 1),
\label{001}
\end{equation}\normalsize
which corresponds to the reduced representation of the usual Pauli-Dirac {\em gamma} matrix representation ($i = 1,\, 2$).
The physical essence of such a theoretical configuration lies in the prediction that fermions can pass through large repulsive potentials without exponential damping.
It corresponds to the so called {\em Klein tunneling} phenomenon \cite{Cal99} which follows accompanied by the production of a particle-antiparticle pair inside the potential barrier.
It is different from the usual tunneling effect since it lies in the energy zone of the Klein paradox \cite{Kle29,Zub80}.
Taking the quadratic form of the above equation reduced to one-dimension for a generic scalar potential $V\bb{x}$, we obtain the analogous Klein-Gordon equation,
\small\begin{equation}
\left(i \partial_{\0} - V\bb{x}\right)^{\2}\phi\bb{k, x} = \left(E - V\bb{x}\right)^{\2}\phi\bb{k, x} = \left(-\partial^{\2}_{\x} + m^{\2}\right)\phi\bb{k, x},
\label{002}
\end{equation}\normalsize
which, from the mathematical point of view, due to the second-order spatial derivatives, has similar boundary conditions to those ones of the Schroedinger equation and leads to stationary wave solutions characterized by a {\em relativistically} modified dispersion relation.

By depicting three potential regions by means of a rectangular potential barrier $V\bb{x}$, $V\bb{x} = V_{\0}$ if $0 \leq x \leq L$, and $V\bb{x} = 0$ if $x < 0$ and $x > L$, we observe that the incident energy can be divided into three zones.
Differently from the energy configuration relative to the non-relativistic (Schroedinger) dynamics, the {\em above barrier} energy zone, $E > V_{\0} + m$, involves diffusion phenomena of oscillatory waves (particles).
In the so called {\em Klein} zone \cite{Kle29,Cal99}, $E < V_{\0} - m$, we find oscillatory solutions (particles and antiparticles) in the barrier region.
In this case, antiparticles see an opposite electrostatic potential to that seen by the particles and hence they will ``see'' a potential well where the particles ``see'' a barrier \cite{Aux1,Kre04}.
The {\em tunneling} zone, $V_{\0} - m < E < V_{\0} + m$, for which only evanescent waves exist \cite{Kre01,Pet03} in the barrier region, is that of interest in this work.

By evaluating the problem for this tunneling (evanescent) zone assuming that $\phi(k,x)$ are stationary wave solutions of the Eq.~(\ref{002}), when the peak of an incident (positive energy) wave packet reach the barrier $x = 0$ at $t = 0$, we can write
\small\begin{equation}
\phi(k,x)=
\left\{\begin{array}{l l l l}
\phi_{\1}(k,x) &=&
\exp{\left[ i \,k \,x\right]} + R(k,L)\exp{\left[ - i \,k \,x \right]}&~~~~x < 0,\\
\phi_{\2}(k,x) &=& \alpha(k)\exp{\left[ - \rho\bb{k}  \,x\right]} + \beta(k)\exp{\left[ \rho\bb{k}  \,x\right]}&~~~~0 < x < L,\\
\phi_{\3}(k,x) &=& T(k,L)\exp{ \left[i \,k (x - L)\right]}&~~~~x > L,
\end{array}\right.
\label{003}
\end{equation}\normalsize
where the novel dispersion relations: $k^{\2} = E^{\2} - m^{\2}$ and $ \rho\bb{k}^{\2} = m^{\2} - (E - V_{\0})^{\2}$, are modified with respect to the usual non-relativistic ones.

In order to proceed with a phenomenological analysis which allows us to establish a correspondence with the non-relativistic (NR) solutions, it is convenient to define the kinematic variables in terms of the following parameters: $w = \sqrt{2 m V_{\0}}$, $\upsilon = V_{\0}/m = w^{\2}/2m^{\2}$, and $n^{\2}\bb{k} = k^{\2}/w^{\2} = E_{NR}/V_{\0}$.
The parameter $w$ corresponds to the same {\em normalization} parameter of the usual NR analysis where $k^{\2} = 2 m E_{NR}$.
The previously quoted relation between the potential energy $V_{\0}$ and the mass $m$ of the incident particle is given by the parameter $\upsilon$.
Finally, $n^{\2}\bb{k}$ represents the dependence on the energy for all the results that are being considered here.
After simple mathematical manipulations, it is easy to demonstrate that the tunneling zone for the above Klein-Gordon equation (\ref{002}) is comprised by the interval $(n^{\2}\bb{k} - \upsilon/2)^{\2} \leq 1$ which made $n^{\2}\bb{k}$ assume larger values ($n^{\2}\bb{k} >> 1$), in opposition to the NR case where the tunneling energy zone is constrained by $0 < n^{\2}\bb{k} < 1$).
We shall observe that such a peculiarity has a suitable relation with the possibility of superluminal transmission through the barrier.
The limits for NR energies ($k^{\2} << m^{\2}$ and $V << m$) given by $\upsilon \,n^{\2}\bb{k} << 1$ and $\upsilon/n^{\2}\bb{k} << 1$ reproduce the Schroedinger equation results for the transmission coefficient and for the corresponding traversal time.

With regard to the {\em standard} one-way direction wave packet tunneling, for the set of stationary wave solutions given by Eq.~(\ref{003}), it is well-known \cite{Ber06} that the transmitted amplitude $T\bb{n, L} = |T\bb{n, L}|\exp{[i \varphi\bb{n, L}]}$ is written in terms of
\small\begin{equation}
|T\bb{n, L}| = \left\{1 + \frac{1}{4 \, n^{\2} \, \rho^{\2}\bb{n}} \sinh^{\2}{\left[\rho\bb{n}\, w L \right]}\right\}^{-\frac{1}{2}},
\label{004}
\end{equation}\normalsize
where we have suppressed from the notation the dependence on $k$, and
\small\begin{equation}
\varphi\bb{n, L} = \arctan{\left\{\frac{n^{\2} - \rho^{\2}\bb{n}}
{2 n \, \rho\bb{n}}
\tanh{\left[\rho\bb{n} \, w L \right]}\right\}},
\label{502}
\end{equation}\normalsize
for which we have made explicit the dependence on the barrier length $L$ (parameter $w L$), and we have rewritten $\rho\bb{k} = w \rho\bb{n}$, with $\rho\bb{n}^{\2} = \sqrt{1 + 2 n^{\2} \upsilon} - (n^{\2} -\upsilon/2)$.

The additional phase $\varphi(n, L)$ that goes with the transmitted wave is utilized for calculating the transit time $t_{\varphi}$ of a transmitted wave packet when its peak emerges at $x = L$,
\small\begin{equation}
t_{\varphi} = \frac{\mbox{d}k}{\mbox{d}E\bb{k}} \frac{\mbox{d}n\bb{k}}{\mbox{d}k} \frac{\mbox{d}\varphi\bb{n, L}}{\mbox{d}n} = \frac{(L)}{v} \frac{1}{w (L)} \frac{\mbox{d}\varphi\bb{n, L}}{\mbox{d}n}
\label{006}
\end{equation}\normalsize
evaluated at $k = k_{\0}$ (the maximum of a generic symmetrical momentum distribution $g(k - k_{\0})$ that composes the {\em incident} wave packet).
By introducing the {\em classical} traversal time defined as $\tau_{\bb{k}} = L (\mbox{d}k/\mbox{d}E\bb{k}) = L / v$, we can obtain the normalized phase time,
\small\begin{equation}
\frac{t_{\varphi}}{\tau_{\bb{k}}}
 = \frac{f\bb{n, L}}{g\bb{n, L}},
\label{007}
\end{equation}\normalsize
\small\begin{eqnarray}
f\bb{n, L} &=&
8 n^{\2} \left[\left(2 + 8 n^{\2} \upsilon + \upsilon^{\2}\right) - \left(4 n^{\2} + 3 \upsilon\right)\sqrt{1 + 2 n^{\2}\upsilon} \right]\nonumber\\
&&~~~~+
4 \left[\left(4 + 4 n^{\2}\upsilon + \upsilon^{\2}\right)\sqrt{1 + 2 n^{\2}\upsilon} - 2 \upsilon \left(2 + 3 n^{\2}\upsilon\right)\right]\frac{\sinh(\rho\bb{n} w L)\,\cosh(\rho\bb{n} w L)}{\rho\bb{n} w L}
\nonumber\\
g\bb{n, L} &=&
16 n^{\2} \left[2 \left(1 + 2 n^{\2}\upsilon\right) -  \sqrt{1 + 2 n^{\2}\upsilon}\left(2 n^{\2} + \upsilon\right)\right]\nonumber\\
&&~~~~+
2 \left[\left(4 + 8 n^{\2}\upsilon + \upsilon^{\2}\right)\sqrt{1 + 2 n^{\2}\upsilon} - 4 \upsilon \left(1 + 2 n^{\2}\upsilon\right)\right]\sinh(\rho\bb{n} w L)^{\2}.
\nonumber
\end{eqnarray}\normalsize

\begin{figure}[th]
\vspace{-0.6 cm}
\centerline{\psfig{file=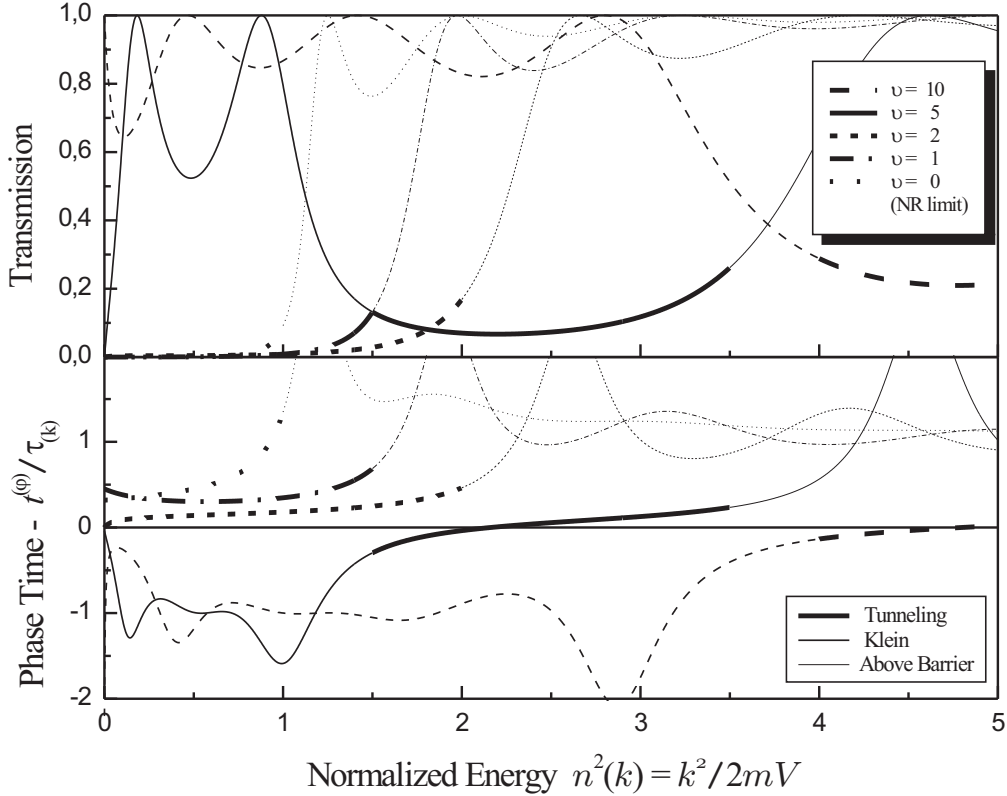,width=15cm}}
\vspace{-1 cm}
\caption{Tunneling TRANSMISSION probabilities and the corresponding tunneling PHASE times for the dynamics of the relativistic wave equation.
We have classified the energy zones by the line thickness: the thick line corresponds to the tunneling energy zone, the intermediate line corresponds to the Klein zone and the thin line corresponds to the above barrier energy zone.
In fact the tunneling region is comprised by the interval intersection $(n^{\2} - \upsilon/2)^{\2} < 1$ and $n^{\2} > 0$.
Here we have adopted the illustratively convenient value of $w L = 2 \pi$.
We have set $\upsilon =  0,\, 1,\, 2 ,\, 5,\, 10$ and we have constrained our analysis to $n^{\2}\bb{k} > 0$ since we have assumed $V_{\0} > 0$.
It is convenient to observe that the NR regime can be parameterized by the limit where $\upsilon \rightarrow 0$.}
\label{Fig01}
\vspace{-0.4 cm}
\end{figure}

We compare the theoretical results for the tunneling phase times in correspondence with their respective transmission probabilities in the Fig.(\ref{Fig01}) for different propagation regimes ($\upsilon = 0 (NR),\, 1,\, 2,\,5,\, 10)$).
It is important to emphasize that the tunneling region is comprised by the interval $(n^{\2} - \upsilon/2)^{\2} < 1, \, n^{\2} > 0$.

We can notice the possibility of accelerated ($t_{\varphi} < \tau_{\bb{k}}$), and eventually {\em superluminal} (negative tunneling delays, $t_{\varphi} < 0$) transmissions without recurring to the usual analysis of the {\em opaque} limit ($\rho\bb{n} w L \rightarrow \infty$) which leads to the Hartman effect \cite{Har62}.
In the NR dynamics (Schroedinger equation solutions), the opaque limit and its consequent superluminal interpretation (Hartman effect) were extensively discussed in the literature.
Superluminal group velocities in connection with quantum (and classical) tunneling were predicted even on the basis of tunneling time definitions more general than the simple Wigner's phase-time \cite{Wig55} (Olkhovsky {\em et al.}, for instance, discuss a simple way of understanding the problem \cite{Olk04}).
Discussions on relativistic causality in addition to several analytical limitations have ruined the most part of possibilities of superluminal interpretation of the NR tunneling phenomena \cite{Lan89,Win03,Ber06}.
In a {\em causal} manner, the arguments consist in explaining the superluminal phenomena during tunneling as simply due to a {\em reshaping} of the pulse, with attenuation, as already attempted (at the classical limit) \cite{Gav84}, i. e. the later parts of an incoming pulse are preferentially attenuated, in such a way that the outcoming peak appears shifted towards earlier times even if it is nothing but a portion of the incident pulse forward tail \cite{Ste93,Lan89}.


We do not intend to extrapolate to the delicate question of whether superluminal group-velocities can sometimes imply superluminal signalling, a controversial subject which has been extensively explored in the literature (\cite{Olk04} and references therein).
Otherwise, the phase time calculation based on the relativistic dynamics introduced here offers distinct theoretical possibilities of  superlunminal transmission in a novel scenario, for the limit case where $\rho\bb{n}\,wL$ tends to $0$ (with $L \neq 0$), in opposition to the opaque limit where $\rho\bb{n} \, wL$ tends to $\infty$.
Let us then separately expand the numerator $f\bb{n, L}$ and the denominator $g\bb{n, L}$ of the Eq.~(\ref{007}) in a power series of $\rho\bb{n}\,wL$ ($\rho\bb{n}  \rightarrow 0$) in order to observe that in the lower (upper) limit of the tunneling energy zone, where
$n^{\2}$ tends to $\upsilon/2 + (-) 1$, the numerical coefficient of the zero order term in $\rho\bb{n}\,wL$ amazingly vanish in the numerator as well as in the denominator!
Since the coefficient of the linear term also is null,
just the coefficient of the second order terms plays a relevant role in both series expansions.
After expanding the Eq.~(\ref{007}), such a {\em step-by-step} mathematical exercise leads to
\small\begin{equation}
\frac{t_{\varphi}}{\tau_{\bb{k}}}
 = \frac{4}{3}
\frac{\left[\left(4 + 4 n^{\2}\upsilon + \upsilon^{\2}\right)\sqrt{1 + 2 n^{\2}\upsilon} - 2 \upsilon \left(2 + 3 n^{\2}\upsilon\right)\right]}{\left[\left(4 + 8 n^{\2}\upsilon + \upsilon^{\2}\right)\sqrt{1 + 2 n^{\2}\upsilon} - 4 \upsilon \left(1 + 2 n^{\2}\upsilon\right)\right]}
+\mathcal{O}(\rho\bb{n}\,wL)^{\2}
\label{0100}
\end{equation}\normalsize
for small values of $\rho\bb{n}$.
At the same time, since $\lim_{n^{\2}\rightarrow \upsilon/2 \pmp 1}{\rho\bb{n}} = 0$, the tunneling transmission probability can be approximated by
\small\begin{equation}
\lim_{n^{\2}\rightarrow \upsilon/2 \pmp 1}{|T\bb{n, L}|} =
\left[1 + \frac{(w L)^{\2}}{2 \upsilon \mp 4}\right]^{-\frac{1}{2}}
\mbox{$\begin{array}{c}\mbox{\tiny$\upsilon >> 1$}\\ \rightarrow \\~\end{array}$}
\left[1 + (m L)^{\2}\right]^{-\frac{1}{2}},
\label{011}
\end{equation}\normalsize
from which we recover the high probability of complete tunneling transmission when $m L << 1$.
Finally, for the corresponding values of the phase times evaluated in (\ref{0100}), we obtain,
\small\begin{equation}
\lim_{n^{\2}\rightarrow \upsilon/2 \mp 1}{\frac{t_{\varphi}}{\tau_{\bb{k}}}} = -\frac{4}{3}\frac{1}{1 \pm 2 n^{\2}}, ~~~~ n^{\2}\rightarrow \upsilon/2 \mp 1, ~~~n^{\2},\,\upsilon > 0,
\label{012}
\end{equation}\normalsize
that does not depend on $m L$, and we notice that its asymptotic (ultrarelativistic) limit always converges to $0$.
In particular, in the lower limit of the tunneling energy zone, $n^{\2}\rightarrow \upsilon/2 - 1$, it is always negative.
Since the result of Eq.~(\ref{012}) is exact, and we have accurately introduced the possibility of obtaining total transmission ({\em transparent barrier}), our result ratifies the possibility of accelerated transmission (positive time values), and consequently superluminal tunneling (negative time values), for relativistic particles when $m L$ is sufficiently smaller than 1 $(\Rightarrow T \approx 1)$.
Keeping in mind that the barrier height has to be chosen such that one remains in the tunneling regime,
it is notorious that the transmission probability depends only weakly on the barrier height, approaching the perfect transparency for very high barriers, in stark contrast to the conventional, non-relativistic tunneling where $T\bb{n, L}$ exponentially  decays with the increasing $V_{\0}$.
Since this results can be analytically extended to the Klein paradox energy zone, such a relativistic effect is usually attributed to a sufficient strong potential that, being repulsive for electrons, is attractive for positrons and results in unstable positron states inside the barrier, which align the energy with the electron continuum outside \cite{Kat06}.

Obviously, the above results correspond to a theoretical prediction, in certain sense, not so far from the experimental realization.
By considering the magnitude of the parameter $m L$ ($m c^{\2}/[\hbar (c/L)]$ in standard units) for an electron with mass $\sim 0.5\,MeV$, and observing that in natural units we have $0.2\, MeV\, pm \sim 1$, we conclude that it should be necessary a potential barrier of width $L << 1\, pm$ to permit the observation of the quoted superluminal transmission.
By principle, its observation makes the effect relevant only for some exotic situations as positron production around super-heavy nuclei ($Z \sim 170$) \cite{Gre85} or evaporation of black holes through generation of particle-antiparticle pairs near the event horizon \cite{Pag05}.

In the most common sense, the above condition should be naturally expected since we are simply assuming that the Compton wavelength ($\hbar/(m c)$) is much larger than the length $L$ of the potential barrier that, in this limit situation, becomes {\em invisible} for the tunneling particle.
The relativistic quantum mechanics establishes that if a wave packet is spread out over a distance $d >> 1/m$, the contribution of momenta $|p| \sim m >> 1/d$ is heavily suppressed, and the negative energy components of the wave packet solution are negligible; the one-particle theory is then consistent.
However, if we want to localize the wave packet in a region of space (wave packet width $d$) smaller than or of the same size as the Compton wavelenght, that is $d < 1/m$, the negative energy solutions (positron states) start to play an appreciable role.
This qualitative arguments report us to the Klein paradox and the creation of particle-antiparticles pairs during the scattering process which might create the intrinsic (polarization) mechanisms for accelerated and/or non-causal fermion teletransportation.
The condition $d < L < 1/m$ (where $d < L$ is not mandatory) imposed over a positive energy component of the incident wave packet in the relativistic tunneling configuration excite the negative energy modes (antiparticles) in the same way that the movement of electrons in a semi-conductor is concatenated with the movement of positively charged {\em holes}.

Turning back to the context of the nanoscopic scale structures, the most challenging possibility of observing similar effects occurs for massless (or effective mass) Dirac fermions in graphene structures.
Even though the linear spectrum of fermions in graphene implies zero rest mass, their cyclotron mass approaches to $ 10^{\mi\2} m_e$ \cite{Nov05}, which increases the superluminal tunneling scale to 1 angstrom.
In spite of the theoretical focus, the results here obtained apply to some configurations which should deserve further attention by experimenters in the study of the graphene structures where the dynamics of the electron is described by a relativistic-like dynamics.
For bi-layer structures, due to the chiral nature of their quasiparticles, quantum tunneling in these materials becomes highly anisotropic, qualitatively different from the case of normal, non-relativistic electrons.
In fact, it has been recently speculated that, from the experimental point of view, the graphene provides an effective medium for mimicking relativistic quantum effects where, for instance, massless Dirac fermions allow a close realization of Klein's gedanken experiment whereas massive chiral fermions in bilayer graphene offer an interesting complementary system that elucidates the basic physics involved.
The point is that, in conventional two-dimensional systems, sufficiently strong disorder results in electronic states that are separated by barriers with exponentially small transmittance \cite{Kat06}.
In contrast, in single- and bi-layer graphene materials all potential barriers are relatively transparent ($T\bb{n, L} \approx 1$): the quasiparticles in graphene exhibit a linear dispersion relation $E = \hbar k v_{f}$ that corresponds to the pseudo-ultrarelativistic limit of our analysis for pseudo-massless particles traveling with their Fermi velocity $v_{f}$.
In this case there are pronounced transmission resonances where $T$ approaches unity for some particular geometric configurations, which does not allow charge carriers to be confined by potential barriers that are smooth on atomic scale.
Moreover, some authors have demonstrated experimentally and theoretically, that the biased graphene bilayer is a tunable semiconductor where the electronic gap can be controlled by the electric field effect reaching values as large as $0.3$ eV \cite{Cas06}.

To summarize, as previously pointed out, we have considered tunneling by a wave packet obeying the Klein-Gordon equation, in a regime where the Compton wavelength of the particle is much larger than the width of the barrier, assumed rectangular.
Our standard analysis have suggested the possibility of total transmission together with an accelerated tunneling.
Once it has been observed that the physical essence of the Klein paradox lies in the prediction that particles can pass through large repulsive potentials without exponential damping \cite{Cal99,Dom99}, the scenario of the Klein-paradox and accelerated tunneling transmission associated with relativistic-like phenomena at nanoscopic scale can be tested experimentally using graphene devices.
In a subsequent study we intend to investigate the appearance of an {\em equivalent} smaller effective mass value $M_{eff} << m$ due the minimal coupling of the charged particle magnetic momentum with an external magnetic field, in particular, for the massless electron propagation in a single-layer graphene which could introduce some novel ingredients for quantifying these peculiarities of the relativistic tunneling effect.

{\bf Acknowledgments}
We thank FAPESP (PD 04/13770-0) for the financial support.

\end{document}